# The influence of individual lattice defects on the domain structure in magnetic antidot lattices

X.K. Hu,[a)] S. Sievers, A. Müller, and H.W. Schumacher

Physikalisch-Technische Bundesanstalt, Bundesallee 100, D-38116 Braunschweig, Germany

Corresponding author:

Xiukun Hu

Physikalisch-Technische Bundesanstalt

Bundesallee 100, 38116

Braunschweig, Germany

Tel: +49(0)531 592 1412

E-mail: xiukun.hu@ptb.de

**ABSTRACT**

We numerically and experimentally investigate the influence of single defects consisting of a missing antidot on the spin configurations in rectangular permalloy antidot lattices. The introduction of such lattice defects leads to the nucleation of complex domain structures after the decay of a saturating magnetic field. Micromagnetic simulations yield four typical domain configurations around the defect having distinct energy densities. The existence of the four spin configurations is confirmed by magnetic force microscopy on antidot lattices containing individual defects.

# I. INTRODUCTION

The modification of the magnetic properties of ferromagnetic thin films by the introduction of antidot lattices has attracted a broad interest over the last years due to promising applications e.g. in the fields of magnetic storage or magnonics.[1-15] The magnetic properties, domain structures, and magnetization reversal properties of various antidot lattices have been investigated as function of lattice parameters such as antidot shape, size, and spacing between the antidots.[2-6] Films with square,[1-2, 5-8] rectangular,[9] and circular antidots,[4, 10-14] having square[2, 4-8, 10-12] and rhombic lattice symmetries[4, 12-14] have been investigated. When such antidot lattices are introduced into a continuous film, spins near the border of the antidots will be strongly pinned, leading to ordered domain structures with the periodicity of the lattice cell. Figure 1(a) sketches the basic equilibrium spin configuration in a rectangular unit cell.[4-7, 10-11] The diagonal equilibrium orientation of the magnetic moment in the cell results in an additional anisotropy[1, 4, 11-12] and thus an enhanced coercivity[6,9] and a modified magnetization reversal mechanism[3, 5-6, 10] compared to the continuous film.

It is a well known fact that defects can strongly influence the magnetization reversal processes of ferromagnetic thin films. They can serve as nucleation centers for domains,[16] or as pinning centers for domain walls.[16-18] However the role of lattice defects such as missing antidots on the spin configurations and magnetization reversal properties of antidot lattices has not been investigated up to now. A possible modification of the spin configuration induced by such defects would be of vital importance for all applications e.g. in the field of magnonics[19]. Here, already a local variation of the spin configuration[20] could significantly modify spin wave propagation in magnonic crystals based on magnetic antidot lattices.

In this paper we investigate numerically and experimentally the influence of lattice defects consisting of missing antidots on the domain structures in rectangular antidot lattices. We find that the introduction of these lattice defects favours the nucleation of so-called super domain walls (SDW), occurring between regions of cells with same average orientation of the magnetic moment, the so-called super domains (SD) of the antidot lattice.[15] Micromagnetic simulations yield four typical domain configurations around the defect having distinct energy densities. The occurrence of these spin configurations is confirmed by magnetic force microscopy (MFM) on antidot lattices containing individual defects.

## II. EXPERIMENTS AND SIMULATIONS

In our studies we consider rectangular antidot lattice geometries as investigated in our previous work.[15] A sketch of an antidot lattice cell is shown in Fig. 1(a). A 30 nm thick permalloy film containing magnetic antidot lattices with circular antidot diameters ($d$=60, 80, and 100 nm) and spacings $S_x$ (100-350 nm), $S_y$ (50-1100 nm) was fabricated by e-beam lithography and lift-off process. The area of every single lattice with the given lattice parameters was 10×10 $\mu m^2$. The domain configurations in the lattices were investigated by a Veeco MFM NanoScope IIIa in a tapping lift mode with high resolution MFM tips (Team Nanotec MFM probe with 25-nm Co alloy coating, 3-N/m spring constant and 75-kHz resonance frequency). MFM images were captured in a remanent state after the decay of an external saturating field of 796 kA/m (10 kOe) in the −$x$ direction. The application of in-plane fields during MFM imaging was not possible using our present setup. In some of the lattices individual lattice defects consisting of missing antidots were found resulting from lithography artifacts. In this work the SD and SDW configurations in such lattices containing defects were experimentally and numerically scrutinized.

Numerical analysis was carried out using a Landau-Lifshitz-Gilbert (LLG) micromagnetic simulator.[21] The computational size was 2400×2400×30 nm$^3$ and 4200×4200×30 nm$^3$, respectively. One antidot in the center of the antidot lattice was left out to create a lattice defect. Periodic boundary conditions in the $x$ and $y$ directions were imposed in order to reduce the influence of the film boundary on the resulting domain configuration. The simulation cell size was 10×10×10 nm$^3$. Typical intrinsic parameters of permalloy were initialized in the simulations as follows: saturation magnetization $M_S$=800 kA/m, exchange stiffness $A$=1.05×10$^{-11}$ J/m and uniaxial anisotropy constant $K_u$=100 J/m$^3$ with easy axis in the $x$ direction.

### III. RESULTS AND DISCUSSION

In previous work,[15] we have studied permalloy films with rectangular antidot lattices with diameter $d \leqslant 100$ nm. In a sufficiently high saturating field along the $-x$ direction, all spins basically point in the $-x$ direction. At a vanishing field afterwards, the average magnetization $m_{av}$ in one cell then can align along two possible diagonal directions with equal probability: either along the [−1, 1] or the [−1, −1] direction. Cells with the same diagonal orientation of $m_{av}$ form a single SD as sketched in Fig. 1(b). At the interfaces between two different SDs with $m_{av}$ in the [−1, 1] and [−1, −1] directions, two types of SDWs are possible, namely SDWs oriented along the $x$ or $y$ orientation, respectively. Fig. 1(c) shows a SDW (black line) along the $x$-axis with a tail-to-tail spin configuration. This type of SDW creates a high divergence of stray field and is referred to as high (stray field) energy SDW (HE-SDW). Also HE-SDWs with a head-to-head spin configuration are possible (not shown). In contrast, Fig. 1(d) shows a SDW (gray line) along $y$-axis with a head-to-tail spin configuration having a low stray field divergence. This type of SDW is referred to as low (stray field) energy SDW (LE-SDW).

First, we numerically analyze the influence of a single defect on the domain structure in a permalloy antidot lattice by micromagnetic simulations. The simulations are performed in lattices with and without a defect during and after a removal of a saturating magnetic field of 238 kA/m (3 kOe) in the $-x$ direction. The results for the lattice parameters $d$=80 nm and $S_x$=$S_y$=150 nm are shown in Fig. 2. It shows a series of snapshots of the micromagnetic simulations of the spin states from full applied saturation field to the remanent equilibrium state at zero applied field. For a defect free antidot lattice the average magnetization $m_{av}$ of all cells uniformly relaxes into the above mentioned lowest energy single SD [Fig. 2 (a-c)]. For the given example $m_{av}$ is oriented along [−1, 1] direction. No LE- or HE-SDWs are present [Fig. 2(c)] in the whole simulated structure.

In contrast, in a lattice with only a single defect a completely different magnetic configuration is found. When the magnetic field is reduced to 79.6 kA/m (1 kOe) $m_{av}$ in the region of the missing antidot marked by the diamond tends to keep the parallel orientation along $-x$ [Fig. 2(d)]. Hence, $m_{av}$ in the region of the surrounding antidots (marked by the dashed square) is deflected from the average orientation in the undisturbed lattice. In the column on the right hand side of the defect $m_{av}$ is deflected from the $-x$ orientation along the applied field towards the missing antidot. In contrast, in the column on the left hand side of the defect $m_{av}$ is deflected in the opposite way, pointing away from the missing antidot. In this way, the stray field energy at the two antidots above and below the defect is reduced. The missing antidot thus leads to the nucleation of four surrounding SDs with different orientation of $m_{av}$ of [−1, 1] (top left and bottom right) and [−1, −1] (top right and bottom left) [Fig. 2(e)]. During further reduction of the field these SDs extend [Fig. 2(f)] resulting in a complex domain pattern at zero field as shown in [Fig. 2(g)].

In Fig. 2(g), a larger section of the simulated area is shown. SDs with [−1, 1] (blue) and [−1, −1] (red) orientations are found, separated by LE-SDWs marked by dashed vertical lines and by a HE-SDWs marked by the horizontal white line. Note again that for the defect free lattice under the same simulation conditions a perfect single SD is formed. This result shows that already the existence of a single individual lattice defect strongly favors the formation of complex domain structures: the defect induces the nucleation of a pair of HE-SDWs around the defect again marked by the dashed rectangular frame in the figure. This spin configuration is most often observed in the simulations as it corresponds to the lowest energy spin configuration. The energy density of this spin configuration calculated by micromagnetic simulations is $7.15 \times 10^3$ J/m$^3$ and thus less than the average energy density of cells within a SD ($8.65 \times 10^3$ J/m$^3$). This is because the parallel spins in the middle of this area reduce the exchange energy density. Furthermore, also the stray field energy is reduced due to the missing antidot.

Similar phenomena are found when lattices with different parameters are simulated. Figure 3 shows the spin configurations of lattices with $d$=80 nm and (a) $S_x$=150 nm, $S_y$=300 nm; (b) $S_x$=300 nm, $S_y$=150 nm; (c) $S_x$=$S_y$=300 nm, and (d) $S_x$=$S_y$=600 nm. For the smaller lattice spacing in Fig. 3(a-c), again a pair of HE-SDWs nucleates around the defect. The defect regions are again marked by dashed rectangular frames. However, the basic domain structure in the cells changes with increasing lattice spacing as the stray field energy is reduced for larger antidot spacings. For example, for the lattice with $S_x$=$S_y$=300 nm (c), the diagonal orientation of $m_{av}$ is less pronounced compared to the lattice with $S_x$=$S_y$=150 nm. As the spacing further increases, the orientation of most of spins is oriented close to the $-x$ direction, and the basic diagonal domain structure disappears. This can be observed in the part marked by a rectangular frame in Fig. 3 (d) for $S_x=S_y=600$ nm. Here, the domain configuration in the area with a defect also changes from a pair of HE-SDW into a single SD. Similar changes of the domain structure are occurring when the antidot diameter decreases from 110 to 50 nm (not shown). The reason for that is that the pinning effect of the antidot can only affect the spins near it but not the whole continuous film area in one cell. Hence an increase of the lattice spacing and a decrease of the antidot diameter lead to similar observations. In both cases the tendency of the defect to induce the nucleation of a HE-SDW is reduced.

In addition to the above discussed spin configuration, the micromagnetic simulations revealed three further characteristic different spin configurations upon variation of the initial simulation conditions. One obvious variation is the introduction of small uniaxial anisotropies of the permalloy film along the *y*-axis or small *y*-components of the initialization field to break the symmetry of the spin initialization along the –*x* orientation. The different possible spin configurations occurring around a defect will be discussed with respect to the lattice with *d*=80 nm and $S_x=S_y=150$ nm. The lowest energy spin configuration in the area around a single defect has been discussed with respect to Fig. 2. Figure 4(a)-(d) shows the four distinct domain configurations found around the missing antidot. The area around the defect is again marked by dashed rectangular frames. The grey scale images shown in the line below are the calculated MFM images resulting from the stray field of the micromagnetic domain configurations.

The aforementioned basic lowest energy spin configuration with a pair of HE-SDWs around the defect is configuration a. It has an energy density of $7.15\times10^3$ J/m$^3$. In the case of configuration b ($7.80\times10^3$ J/m$^3$), the average magnetization vectors in this area all point diagonally up, forming one single SD. In configuration c ($8.28\times10^3$

J/m$^3$), SDs with an average magnetization vector pointing diagonally up and down are found on the left and right hand side of the defect, respectively. A LE-SDW is thus formed in this area. Configuration d (9.42×10$^3$ J/m$^3$) consists of one columnar SD on the left hand side and one head-to-head HE-SDW on the right hand side of the defect. The MFM signals calculated for the four different domain configurations around the defect are well distinct. Hence it should be possible to distinguish such different domain configurations in MFM measurements on individual defects.

Our MFM experiments were carried out on antidot lattices with a stochastic number of unintentional defects resulting from lithography artifacts. Therefore systematic studies of a considerable number of defects were only possible on a very limited set of antidot arrays. The largest numbers of defects revealing distinct domain structures around them were found in antidot lattices with antidot diameter $d$=100 nm and lattice spacings of $S_x$=$S_y$=300 nm. Here five defects were found in the lattice. Figure 5 shows an atomic force microscope measurement (a) and three MFM images (b-d) of this lattice. The five defects are marked by arrows. Like in the simulations, the sample was saturated in an applied magnetic field along the –$x$ direction prior to the MFM measurements. The MFM images were then taken in the remanent state at vanishing field. This saturation and measurement cycle was repeated a few times in order to observe different spin configurations around the defects resulting from different stochastic nucleation processes and a small unintentional stray ($\leq \pm 2°$) of the in-plane orientation of the saturation field. In the MFM images of Fig. 5 (b)-(d) the signature of all four computed domain configurations is experimentally observed. In the figures these domain configurations are marked by the arrows a-d. Figure 5(b) shows area with all the five defects showing the domain configurations a, b, and c (arrows). Figure 5(c) and 5(d) show the inner part of the same area with the three central defects (arrows). The three MFM images were taken after different saturation cycles. In one of the measurements in Fig. 5(c) also the configuration d is present. Enlarged MFM images of the four domain configurations a-d can also be found in Fig. 6(a-d) (top panel).

As shown above different domain configurations can be found after different saturation cycles underlining the importance of stochastic magnetization processes during domain nucleation. Furthermore the slight symmetry breaking of the angular stray of the saturation field might play a role on the nucleation processes. In the series of three MFM measurements of Fig. 5 the configuration b was found nine times. It hence seems to be the most probably configuration. The reason for this might be related to the following: only the configuration b does not induce HE-SDWs in the antidot lattice. Therefore the lattice can keep a broad single SD which reduces the overall energy. While the configuration a is found four times and configurations c and d are only found once.

As mentioned before a broad range of antidot lattices with different lattice spacings ($S_x$, $S_y$) have been prepared and investigated. However, only few of them contained defects resulting from lithography artifacts. The lattices constants of these lattices were $d = 100$ nm, and ($S_x$, $S_y$) = (350 nm, 250 nm), (300 nm, 250 nm) and (250 nm, 250 nm). Though these lattices have been investigated by MFM (not shown) the small number of defects in these lattices (between one and three) inhibits a systematic study of the occurring defect spin configurations as function of lattice parameters. Note however that also in these lattices the spin configuration b seems to be the predominant configuration.

In the following we will discuss the experimentally observed spin configurations for the lattice with $d=100$ nm, and $S_x=S_y=300$ nm (cp. Fig. 5) by comparison to micromagnetic simulations. In Fig. 6 detailed MFM images (a)-(d) of the four occurring spin configurations a-d are displayed in the top panel from left to right, respectively. The middle panel (e)-(h) shows the corresponding simulated domain configurations, and the bottom panel shows the corresponding computed MFM images which is based on the simulation data.

These simulations with slightly different initialization parameters as discussed above often reveal the same two basic configurations a and b as shown in Fig. 6(e) and (f). The corresponding energy densities are $3.15 \times 10^3$ J/m$^3$ (a), and $2.91 \times 10^3$ J/m$^3$ (b). The characteristics of these two spin configurations are basically very similar to those in the lattice with $d$=80 nm and $S_x$=$S_y$=150 nm as discussed with respect to Fig. 4. Here the main difference is the somewhat less contrasted MFM images due to the reduced stray fields occurring for larger antidot separation. In contrast to simulations of the lattice with smaller spacing, the two higher energy configurations c and d have not been found in our simulations. Here the simulations hence do not well agree with the experimental results. Note however that our simulations are based on a model of an ideal antidot lattice with perfect regular spacing and circular antidots. Here, irregularities of the lithography of the different antidots of the lattice as well as a locally varying intrinsic anisotropy of the permalloy thin film might lead to additional symmetry breaking and hence to the nucleation of the two other spin configurations c and d in the lattice. However at present the reasons for the discrepancy of simulation and MFM measurements for the given lattice parameters could only be subject of speculation.

Note that in the simulations it was possible to observe the two remaining spin configurations c and d when introducing a local pinning of spins in the vicinity of the defect. This could be achieved e.g. by locally disturbing the spin configuration in the vicinity of the missing antidot by introducing additional lattice defects only few lattice constants away from the defect under investigation. The resulting spin configurations and computed MFM images are shown in panels (g) and (h) of Fig. 6. For all resulting simulated defect spin configurations (e)-(h) the computed MFM images well capture the main features of the experimental data.

The energy densities of these extrinsically stabilized spin configurations c and d are $2.44\times10^3$ J/m$^3$ and $2.51\times10^3$ J/m$^3$, respectively. Both spin configurations thus exhibit lower energy densities than configurations a and b. This is astonishing since they were not observed in simulations without introducing additional defects. This discrepancy may be related to the weaker pinning of the spins near the antidots for the larger lattice spacings. In the lattice with small spacings of Fig. 4 ($d$=80 nm and $S_x$=$S_y$=150 nm) the boundary of the antidots can affect the whole cell due to the small continuous film area in the cell. The spins between two nearest antidots are hence strongly pinned which restricts the orientation of $m_{av}$ in the cells. Therefore, configurations c and d can be formed in the lattice with $d$=80 nm and $S_x$=$S_y$=150 nm as metastable configurations although they exhibit a higher energy density.

In contrast, the antidot has a relatively small influence on the domain structure in the lattice with larger spacing because of a large continuous permalloy area in one cell compared to the antidot. The pinning effect of the antidot boundary thus mainly influences the spins near the antidot. Therefore the spins in the area of the missing antidot first all align along the same direction when the saturation field is reduced. This way a low exchange energy density state in this large area is achieved. This parallel spin configuration in the defect area, characteristic of the two spin configurations a and b, explains the occurrence of these two configurations and prevents the formation of configurations c and d. However, with additional pinning of nearby spins (e.g. by the introduction of additional nearby defects) this effect can be overcome and the two other spin configurations could be nucleated and observed in the simulations.

## IV. CONCLUSIONS

We have investigated the influence of individual lattice defects consisting of a missing antidot on the spin configurations in rectangular permalloy antidot lattices. Micromagnetic simulations show that already a single lattice defect can act as nucleation centre for complex domain structures. The occurrence of the predicted micromagnetic domain configurations has been confirmed by MFM. Such individual defects should have a strong impact on the magnetization reversal properties and might give a new handle to tailor the magnetization reversal processes in antidot lattices or to obtain specific local magnonic properties in such structures.

## ACKNOWLEDGEMENTS

The research was performed within the EMRP JRP IND 08 MetMags. The EMRP is jointly funded by the EMRP participating countries within EURAMET and the EU.

**FIGURE CAPTIONS:**

FIG. 1 (a) Sketch of the basic spin configuration in one cell. (b) Lattice containing a single super domain (SD) with $m_{av}$ oriented along [−1, 1]. (c) Lattice containing two SDs separated by a high energy super domain wall (HE-SDW; black line). (d) Lattice containing two SDs separated by a low energy super domain wall (LE-SDW; gray line).

FIG. 2 Snapshots of micromagnetic simulations of the development of the magnetic spins from saturation to zero field equilibrium during and after removal of the saturating magnetic field. (a-c) shows lattice without and (d-g) with a defect. Lattice parameters are $d$=80 nm and $S_x$=$S_y$=150 nm. Dashed lines in (g) indicate LE-SDWs. The white horizontal lines in (g) indicate HE-SDWs.

FIG. 3 Simulated spin configurations around one defect in the lattices with $d$=80 nm and (a) $S_x$=150, $S_y$=300 nm; (b) $S_x$=300, $S_y$=150 nm; (c) $S_x$= $S_y$=300 nm and (d) $S_x$= $S_y$=600 nm.

FIG. 4 (a)-(d) Top: Micromagnetic configurations of the four different simulated spin configurations in the area around a defect. Lattice parameters are $d$=80 nm and $S_x$=$S_y$=150 nm. (a)-(d) Bottom: calculated MFM images for the above simulated magnetic configurations.

FIG. 5 (a) Topography and (b-d) MFM images measured on a lattice with $d$=100 nm and $S_x$=$S_y$=300 nm, respectively. Arrows indicate the positions of defects. The letters in (b-d) indicate the four spin configurations around the defects corresponding to the simulated configurations of Fig. 4(a)-(d).

FIG. 6 (a-d) Enlarged MFM images of the four spin configurations observed experimentally in the lattice with $d$=100 nm and $S_x$=$S_y$=300 nm. Middle panel (e)-(h): Simulated spin configurations corresponding to the experimental data. Bottom: calculated MFM images for the above simulated magnetic configurations.

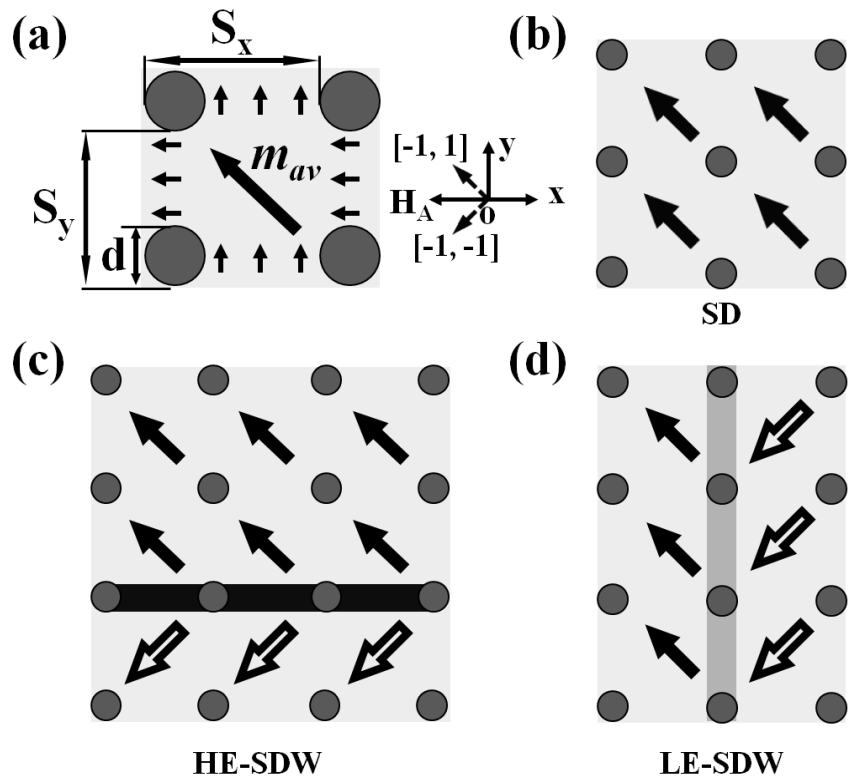

FIG. 1

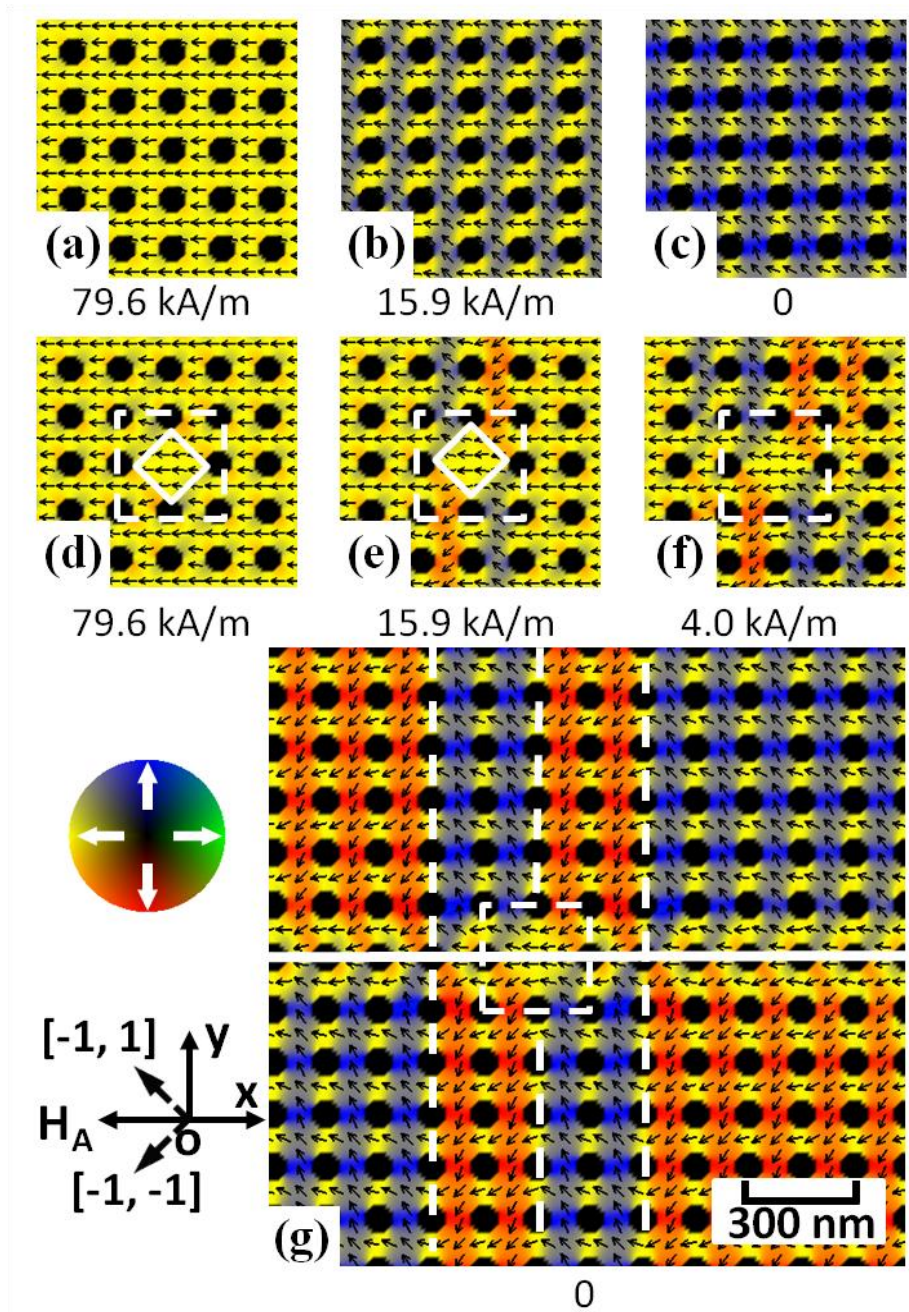

FIG. 2

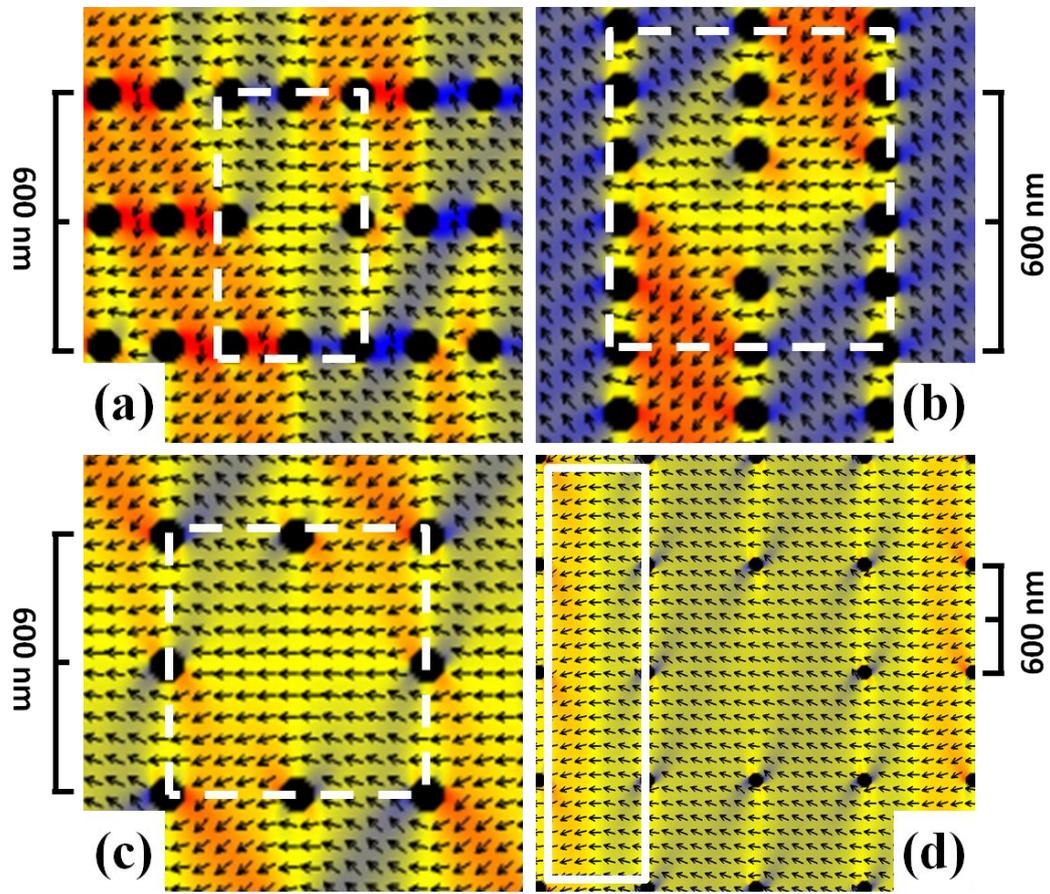

FIG. 3

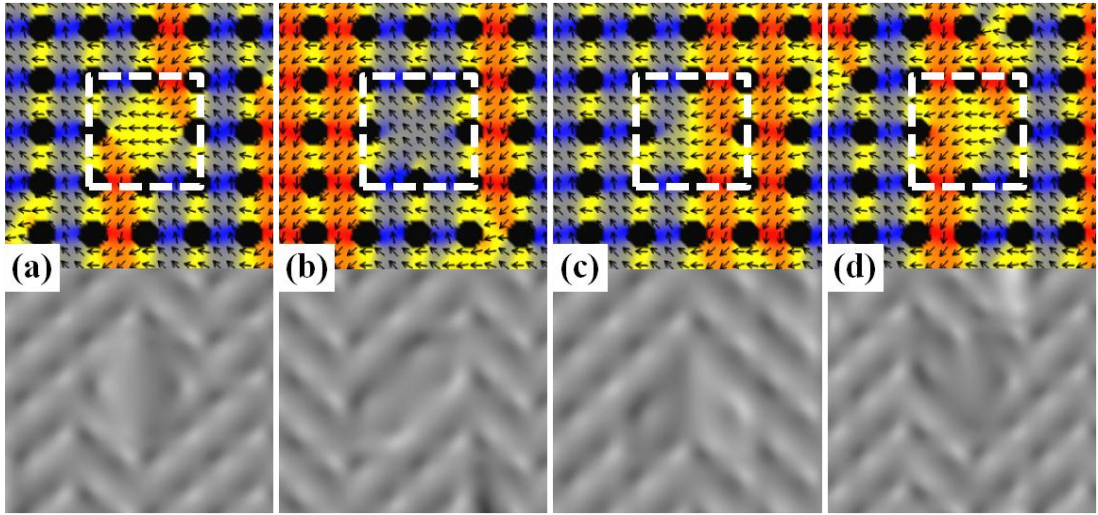

FIG. 4

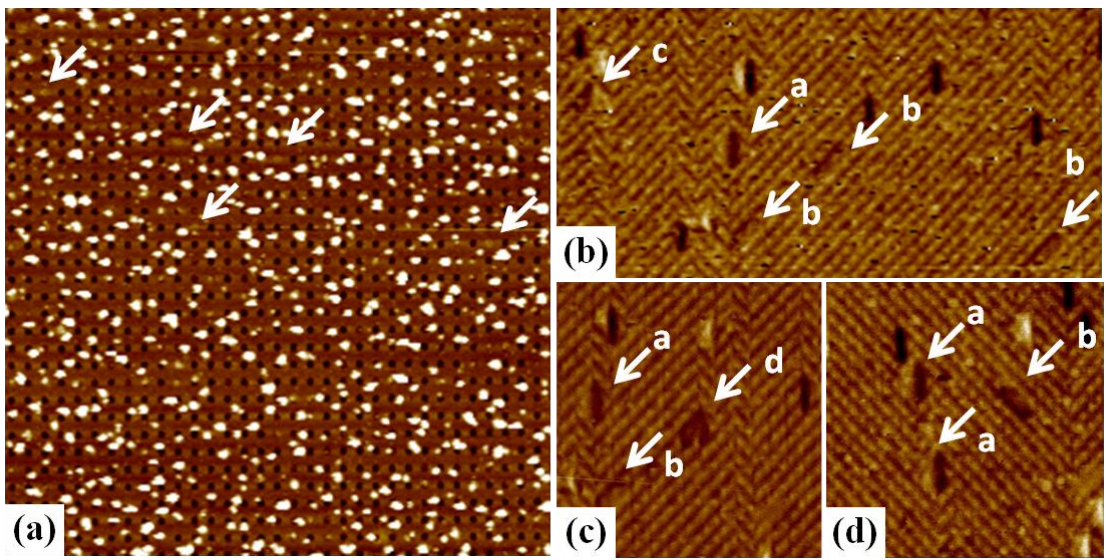

FIG. 5

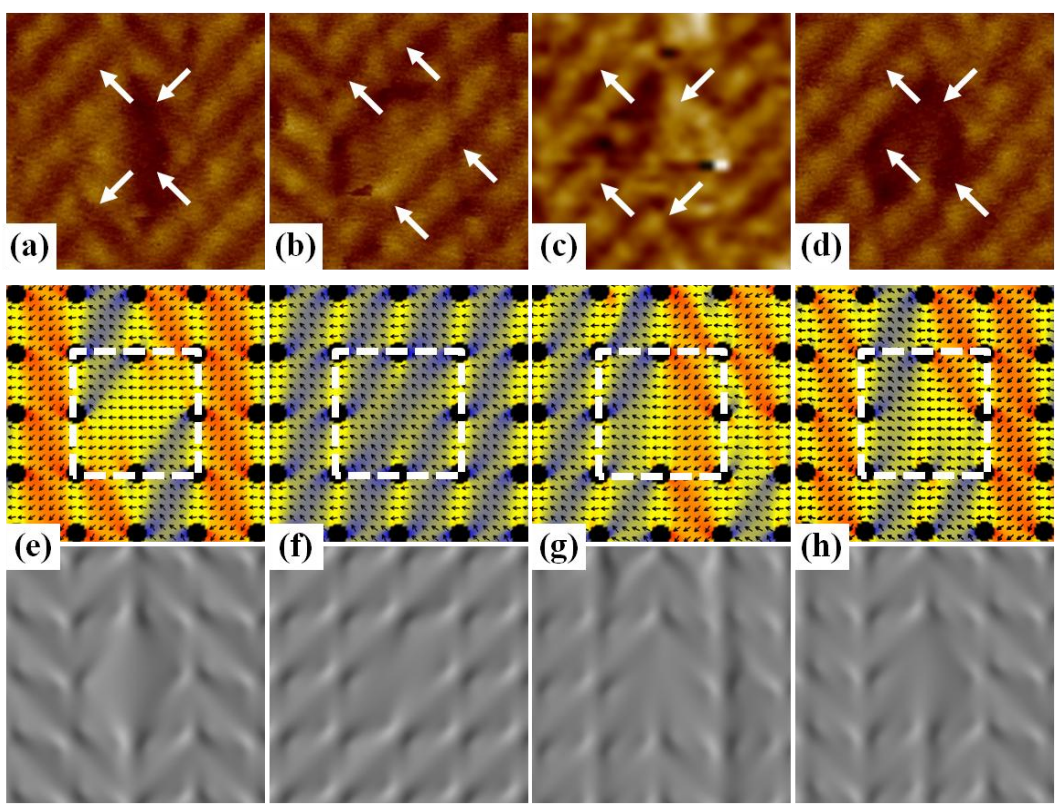

FIG. 6